\documentclass[12pt,a4paper]{article}

\usepackage[british]{babel}

\usepackage[a4paper,top=2cm,bottom=2cm,left=2.5cm,right=2.5cm,marginparwidth=1.75cm]{geometry}
\usepackage{mystyle}

\title{Recursive Interferometric Surface-wave Suppression For Improved Reflection Imaging}

\author[1]{Faezeh Shirmohammadi}
\author[1*]{Deyan Draganov}
\author[1]{Ranajit Ghose}
\author[1]{Eric Verschuur}
\author[1]{Jan Thorbecke}
\author[1]{Kees Wapenaar}

\affil[1]{\small Department of Geoscience and Engineering, Delft University of Technology, 2628 CN Delft, The Netherlands}
\affil[*]{Corresponding author: \texttt{D.S.Draganov@tudelft.nl}}

\date{}  
\begin{document}
\maketitle

\begin{abstract}
\onehalfspacing
High-resolution seismic reflections are essential for imaging and monitoring applications. In seismic land surveys using sources and receivers at the surface, surface waves often dominate, masking the reflections.
In this study, we demonstrate the efficacy of a two-step procedure to suppress surface waves in an active-source reflection seismic dataset. First, we apply seismic interferometry (SI) by cross-correlation, turning receivers into virtual sources to estimate the dominant surface waves. Then, we perform adaptive subtraction to minimise the difference between the surface waves in the original data and the result of SI. We propose a new approach where the initial suppression results are used for further iterations, followed by adaptive subtraction. This technique aims to enhance the efficacy of data-driven surface-wave suppression through an iterative process.
We use a 2D seismic reflection dataset from Scheemda, situated in the Groningen province of the Netherlands, to illustrate the technique's efficiency. A comparison between the data after recursive interferometric surface-wave suppression and the original data across time and frequency-wavenumber domains shows significant suppression of the surface waves, enhancing visualization of the reflections for following subsurface imaging and monitoring studies. \par
\vspace{\baselineskip}
\end{abstract}
  
\newpage
\section{Introduction}
Seismic data acquired on land is often contaminated by surface waves, which are a significant noise source for reflection seismic studies. The surface waves often have a velocity and frequency content similar to those of the investigated reflections, which makes it challenging to suppress them from the dataset. 
Conventionally, surface waves are suppressed during data processing using methods such as frequency-offset (f-x) (\cite{Yilmaz2001}), frequency-wavenumber (f-k), or bandpass filtering. However, these approaches can prove ineffective when surface waves are scattered and overlap with the frequency and moveout of the reflected body waves that we intend to preserve. The f-k filter may result in artefacts due to signal distortion and spatial correlation of the background noise because the surface-wave energy may be distributed over a significant range of the f-k spectrum (\cite{KONSTANTAKI201528}), thus further lowering the quality of the reflections. \par
Recently, the prediction of surface waves with seismic interferometry (SI) and their subsequent adaptive subtraction from seismic reflection data has emerged as a technique for suppressing surface waves (\cite{Dong2006}; \cite{Halliday2010}). In SI, seismic observations from various receiver locations are, for example, cross-correlated to retrieve new seismic responses from virtual sources positioned at the receivers' locations (\cite{Lobkis}; \cite{CampilloandPaual2003};  \cite{Wapenaar2006GreensInterferometry}; \cite{vanmanen2006}; \cite{Curtis2006}). This process enables the retrieval of seismic responses between pairs of receivers. For suppression, the retrieved responses are then subtracted from the original field recordings using least-squares matching, resulting in data with suppressed surface waves. This suppression technique is usually called interferometric surface-wave suppression. \par
In previous studies, the interferometric surface-wave suppression was applied in a non-recursive way on the data. \textcite{Halliday2010} demonstrated its effectiveness in the context of hydrocarbon exploration, while \textcite{KONSTANTAKI201528} and \textcite{Liu2018SeismicWaves} showcased its utility for near-surface applications. Moreover, \textcite{Balestrini2020ImprovedSuppression} demonstrated its application for deep mineral explorations. Here, we propose a new approach using the first output of the interferometric surface-wave suppression for more iterations. We term this technique "Recursive Interferometric Surface-wave Suppression" (RISS). This technique aims to enhance the efficacy of the data-driven surface-wave suppression through an iterative procedure. \par 
In this study, we demonstrate RISS on a 2D reflection dataset acquired in Scheemda, Groningen province, the Netherlands. By using RISS, we aim to enhance the visualisation of reflections, which can provide clearer images of the subsurface structures and enhance the overall interpretation of the seismic data. Such advancements are particularly critical for Groningen, where gas production has resulted in induced seismicity since 1963 (\cite{Muntendam-Bos2022AnNetherlands}). We evaluate the RISS results in comparison with those from other techniques such as time muting and f-k filtering.\par
Below, we first present in Section 2 the methodology of RISS. This will be followed by a description of the seismic data acquisition in Section 3, the results in Section 4, and then a discussion and conclusions. \par

\section{Methodology}\label{Method}
In our proposed approach, SI is employed first to retrieve the dominant surface waves. The retrieved surface-wave energy is subsequently adaptively subtracted from the dataset. Following this, the obtained data is utilised to iterate through these two steps, contributing to the improvement of the reflection resolution. This section outlines the implementation of RISS.
\subsection{Surface-wave retrieval by seismic interferometry}
SI refers to the method of retrieving new seismic responses, for example between two receivers, using most commonly cross-correlation, and the result creates a virtual source at one of the receiver locations (\cite{Wapenaar2006GreensInterferometry}; \cite{Larose2006GEO}; \cite{Schuster2006GEO}). In an active-source survey, this process is usually achieved by cross-correlating the recordings at two receivers and then stacking the individual virtual-source traces over all available active sources (\cite{Halliday2008}). So, the retrieved virtual-source response between two receivers at positions $\mathbf{x}_{A}$ and $\mathbf{x}_{B}$ can be expressed in the time domain in its simplest form as:
\begin{equation}
   G(\mathbf{x}_{B},\mathbf{x}_{A},t) + G(\mathbf{x}_{B},\mathbf{x}_{A},-t) = \sum_{n=1}^{N}
   G(\mathbf{x}_{B},\mathbf{x}_{n},t) * G(\mathbf{x}_{A},\mathbf{x}_{n},-t) ,
   \label{eq:5_1}
\end{equation}

\noindent where $G(\mathbf{x}_{B},\mathbf{x}_{n},t)$ is the response of a recording at receiver $\mathbf{x}_{B}$ and $G(\mathbf{x}_{A},\mathbf{x}_{n},-t)$ is the time-reversed response of a recording at receiver $\mathbf{x}_{A}$, both from a source at $\mathbf{x}_{n}$ at the Earth's surface. The left-hand side of the equation represents the response and its time-reversal between the two receivers at $\mathbf{x}_{A}$ and $\mathbf{x}_{B}$ at the surface, implying that the receiver at $\mathbf{x}_{A}$ has been turned into a virtual source. $N$ represents the total number of active sources at the surface and $*$ denotes convolution.\par
In a laterally homogeneous 2D medium, sources at points in line with the receivers contribute to the retrieval of direct surface-wave arrivals since they are all in the stationary-phase region. So, the results retrieved by SI will be dominated by surface waves (\cite{Balestrini2020ImprovedSuppression}).\par
\begin{figure}[ht]
    \centering
    \includegraphics[width=0.7\textwidth]{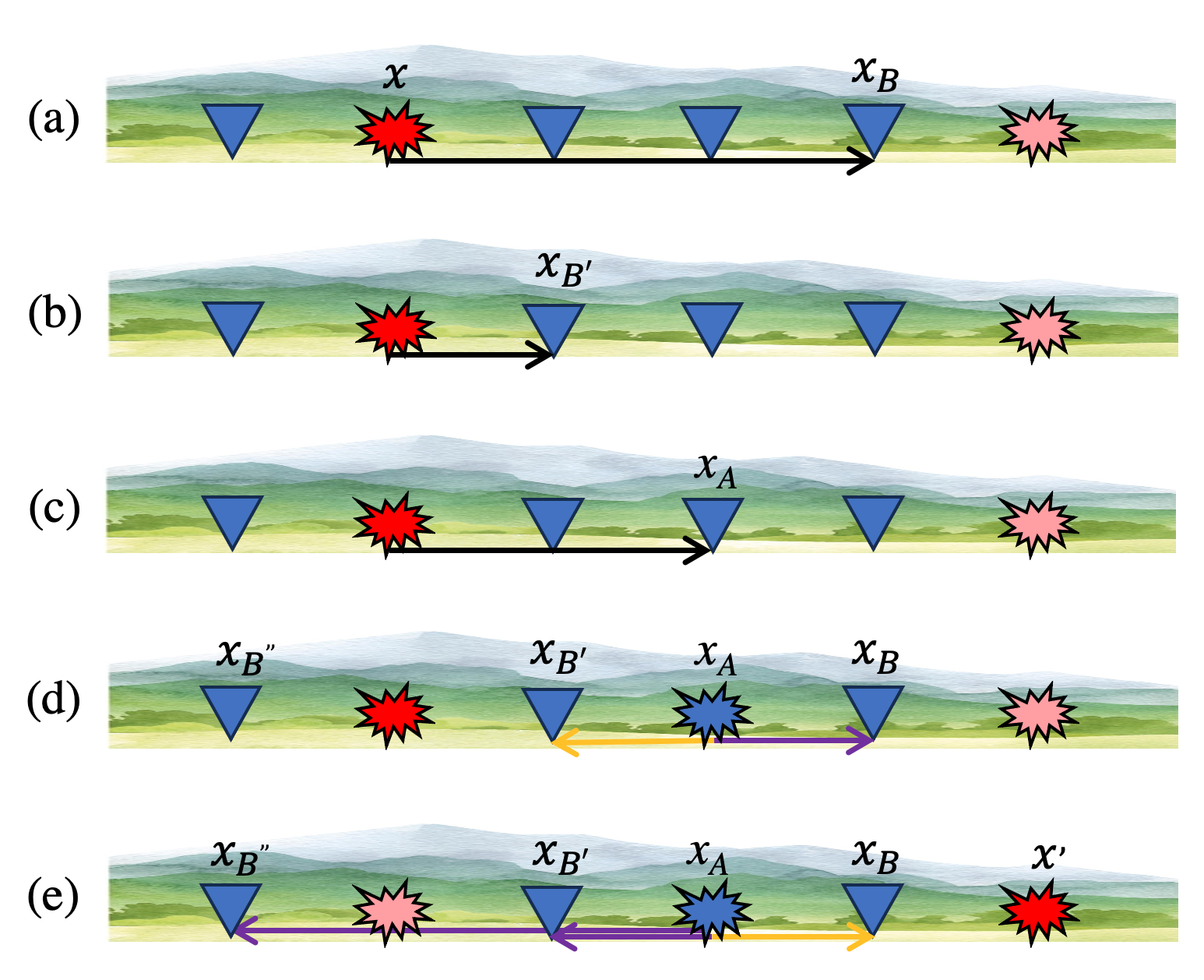} 
    \caption{Schematic representation of seismic interferometry for retrieving surface waves. (a), (b), and (c) The surface wave from the active source $\mathbf{x}$ recorded at $\mathbf{x}_{B}$, $\mathbf{x}_{B'}$, and $\mathbf{x}_{A}$, respectively. (d) The results of correlating the response at $\mathbf{x}_{A}$ with those at other receivers, and thus turning $\mathbf{x}_{A}$ into a virtual source. (e) Same as (d) but for the active source at $\mathbf{x}^{'}$. The black arrows indicate the surface waves, while the orange and purple arrows represent the results of correlation, considering the causal and acausal parts, respectively.}
    \label{fig:5_1}
\end{figure}

Figure~\ref{fig:5_1} shows a schematic representation of SI for retrieving direct arrivals, including surface waves. By correlating the recording at $\mathbf{x}_{B}$ from the active source at $\mathbf{x}$ in Figure~\ref{fig:5_1}a with a recording at $\mathbf{x}_{A}$ in Figure~\ref{fig:5_1}c, the virtual response between $\mathbf{x}_{B}$ and $\mathbf{x}_{A}$ is retrieved, as illustrated by the purple arrow in Figure~\ref{fig:5_1}d, at causal times (the causal part refers to times later than the zero time). Similarly, the virtual response between another receiver at $\mathbf{x}_{B'}$ and a receiver at $\mathbf{x}_{A}$ is retrieved by correlating the response at $\mathbf{x}_{B'}$ in Figure~\ref{fig:5_1}b with that at $\mathbf{x}_{A}$ in Figure~\ref{fig:5_1}c, as depicted by the orange arrow in Figure~\ref{fig:5_1}d, at acausal times (the acausal part refers to times earlier than the zero time). In both cases, the receiver at $\mathbf{x}_{A}$ acts as a virtual source as shown by the blue explosion in Figure~\ref{fig:5_1}d. We repeat this procedure for all active sources, e.g., as shown in Figure~\ref{fig:5_1}e for another active source at $\mathbf{x}^{'}$. Finally, the Green's function and its time-reversal between the virtual source at $\mathbf{x}_{A}$ and other receivers at $\mathbf{x}_{B}$, $\mathbf{x}_{B'}$, and  $\mathbf{x}_{B''}$ are retrieved by stacking all virtual responses such as those shown in Figures~\ref{fig:5_1}d and~\ref{fig:5_1}e. \par
When we want to apply this technique to a field dataset, there are certain issues that need to be addressed in order to improve the resolution of the retrieved responses. First, we aim to retrieve the surface waves with SI. So, it is required that all receivers be considered on the same side of the active source, e.g, for an active source at $\mathbf{x}$ in Figure~\ref{fig:5_1}a, we correlate the response for receivers $\mathbf{x}_{B}$ and $\mathbf{x}_{B'}$ located on the same side as the virtual source at $\mathbf{x}_{A}$, as shown in Figures~\ref{fig:5_1}a, b, and c. In the same way, for an active source at $\mathbf{x}^{'}$ in Figure~\ref{fig:5_1}e, we consider all receivers because they are on the same side as the virtual source at $\mathbf{x}_{A}$, as shown in Figure~\ref{fig:5_1}e. This condition is dictated by the theory of SI by cross-correlation, which states that the sources should surround the receivers, i.e., there must be no sources located between the receivers involved in the correlation process. \par
Second, in the case of isotropic illumination of the receivers, a time-symmetric response between the receivers is obtained, as shown in equation~\ref{eq:5_1}. Consequently, one could sum the causal and the time-reversed acausal parts of the correlated panels to improve the signal-to-noise ratio. However, in practical situations, when the illumination is not homogeneous from all sides for each pair of receivers, then parts of the response can be retrieved at acausal times and other parts at causal times. Therefore, to enhance the quality of our retrieved responses, we meticulously assess the positions of virtual source-receiver pairs and active sources. Subsequently, we opt to select either the causal or time-reversed acausal part of the correlation panel. \par
Considering the conditions of one-sided distribution of receivers and causality, we limit ourselves to a minimum number of traces for stacking. To maintain a high signal-to-noise ratio, we stack traces only when we have at least half the number of all active sources. We can summarise this as follows:
\begin{equation*}
    trace
    \left\{
    \begin{array}{ll}
       \text{causal part, } \text{if }  m>N/2  \text{ with }P_\mathbf{x} < P_{\mathbf{x}_{A}} < P_{\mathbf{x}_{B}} \text{, }\text{or } P_\mathbf{x} > P_{\mathbf{x}_{A}} > P_{\mathbf{x}_{B}}, \text{or } P_{\mathbf{x}_{A}} = P_{\mathbf{x}_{B}} \\
        \text{time-reversed acausal, }   \text{if } m>N/2  \text{ with } P_{\mathbf{x}} < P_{\mathbf{x}_{B}} < P_{\mathbf{x}_{A}} \text{, }\text{or } P_{\mathbf{x}} > P_{\mathbf{x}_{B}} > P_{\mathbf{x}_{A}}.
    \end{array}
    \right.
\end{equation*}

Here, the scalar $P_\mathbf{x}$ is the position of active sources along a 2D seismic line, $P_{\mathbf{x}_{A}}$ is the position of the virtual source, $P_{\mathbf{x}_{B}}$ is the position of the receiver, and $m$ is the number of active sources for stacking, which should be greater than half the total number of active sources ($N$). For instance, as illustrated in Figure~\ref{fig:5_1}d, for the receiver at $\mathbf{x}_{B'}$ on the left side of the virtual source at $\mathbf{x}_{A}$, we consider the acausal part, as indicated by the orange arrow. Similarly, for a receiver at $\mathbf{x}_{B}$ on the right side of the virtual source, we consider the causal part, indicated by the purple arrow. Furthermore, as depicted in Figure~\ref{fig:5_1}e, for the receiver at $\mathbf{x}_{B}$ on the right side of the virtual source at $\mathbf{x}_{A}$, we consider the acausal part, and for other receivers at $\mathbf{x}_{B'}$ and $\mathbf{x}_{B''}$, we consider the causal part, as shown by the purple arrows. Considering the above factors, we retrieve the virtual common-source gather for all receivers. \par

\par
\subsection{Adaptive subtraction}
When each source position in an active-source survey is in close proximity to a receiver position, we are able to identify a corresponding retrieved virtual common-source gather with estimated dominant surface waves for each active source–virtual source pair. These estimates can then be adaptively subtracted from the complete responses of the active sources (\cite{Halliday2008}, \cite{Halliday2010}, \cite{KONSTANTAKI201528}).
To perform adaptive subtraction, we estimate a shaping filter \textbf{f} that can minimise the following objective function (\cite{Liu2018SeismicWaves}; \cite{Balestrini2020ImprovedSuppression}):
\begin{equation}
   \displaystyle \min_{\boldsymbol{f}} \lVert {D} - \boldsymbol{f}{D}_{\text{SW}}\rVert^2 ,
   \label{eq:5_2}
\end{equation} 

\noindent where $D$ stands for the field dataset with surface waves and ${D}_{SW}$ stands for the surface waves retrieved by SI using the field dataset. The squared vertical double bars $||.||^2$ represent the L2 norm. Equation~\ref{eq:5_2} is solved using an iterative least-squares fit (\cite{Verschuur1992}). More details can be found in \textcite{Alai} and \textcite{seismic-multiple-removal}. The product of estimated \textbf{f} and ${D}_{SW}$ is directly subtracted from $D$, giving ${D_{ref}}$ which represents the data after surface-wave suppression as \par
\begin{equation}
   {D_{\text{ref}}} = {D} - \boldsymbol{f}{D}_{\text{SW}}.
   \label{eq:5_3}
\end{equation}

The data after adaptive subtraction may still contain surface waves due to, for example, errors in estimating higher modes of surface waves. Therefore, we suggest repeating the same step of SI and adaptive subtraction but now using the output of the first adaptive subtraction as input for SI. So, we estimate surface waves from SI, and then adaptively subtract them from the output of the first iteration. Repeating these steps improves our chances of suppressing surface-wave energy, as demonstrated in the numerical example in Appendix A, thereby enhancing the resolution of reflections. We call this technique RISS. Note that RISS can be applied for one iteration or multiple iterations.

\section{Seismic data acquisition}
\begin{figure}[H]
    \centering
    \includegraphics[width=0.9\textwidth]{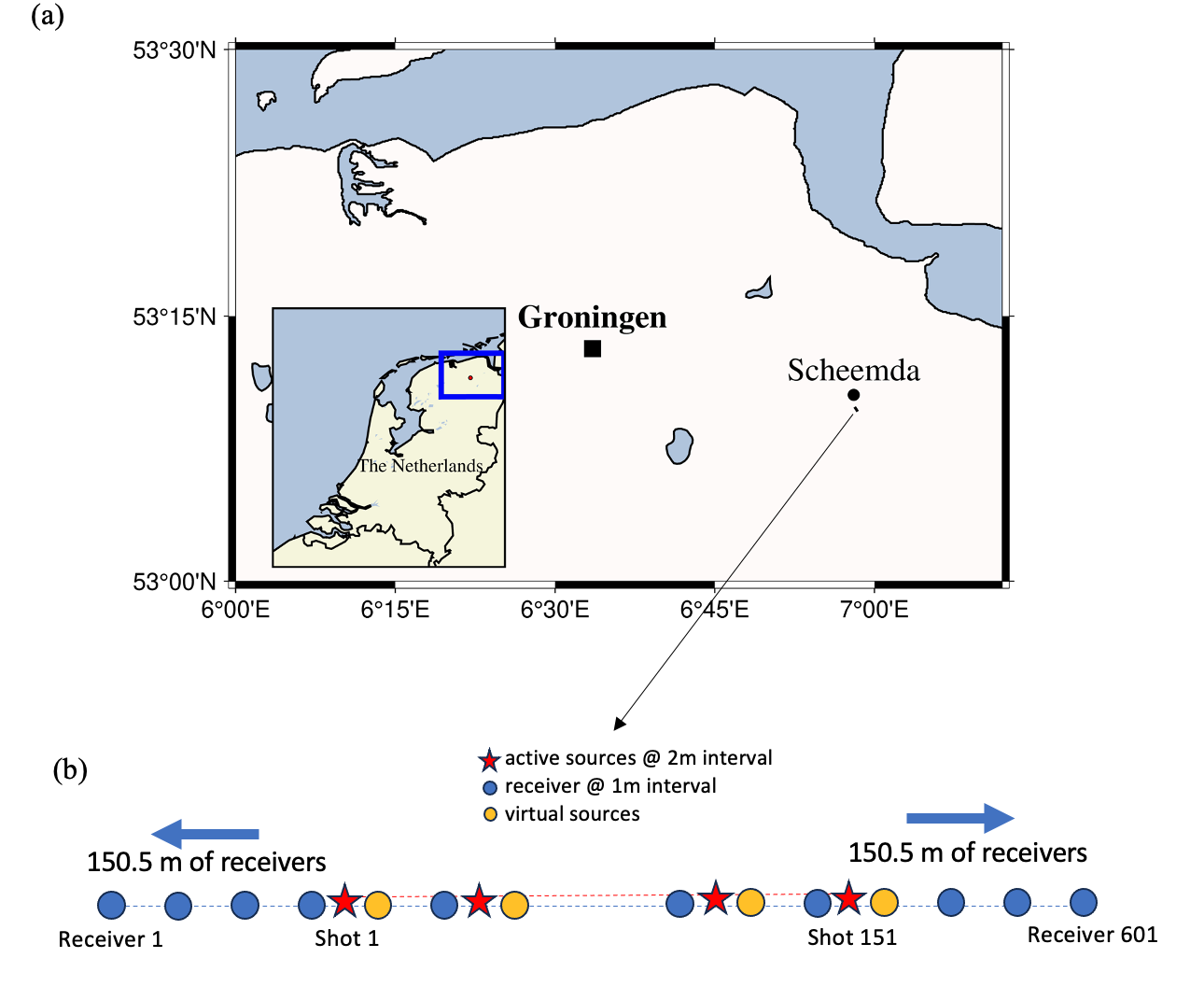} 
    \caption{(a) Location of the test site, and (b) the geometry of the seismic line. The red stars represent active sources, the blue circles represent receivers, and the orange circles represent receivers which act as virtual sources for applying the RISS.}
    \label{fig:5_3}
\end{figure}

We acquired a 2D seismic reflection dataset close to the town of Scheemda in the Groningen province of the Netherlands in 2022. Figures~\ref{fig:5_3}a and ~\ref{fig:5_3}b show the location of the site and the geometry of the reflection line, respectively. We employed an electrical linear motor system (LMS) seismic vibrator (\cite{Noorlandt2015AMotors}) as a source, with a spacing of 2 m (red stars in Figure~\ref{fig:5_3}b), and 601 three-component geophone nodes as receivers (the circles in Figure~\ref{fig:5_3}b), with a spacing of 1 m. The acquisition parameters are summarised in Table~\ref{tab:5_1}.\par
For this survey, we used the electrical vibrator in the S-wave mode and oriented it in the crossline direction. We then used the data recorded by the crossline horizontal component of the geophones. Because of the orientation of the sources and the receivers, and assuming no scattering from the crossline direction, the horizontally polarised S-waves (SH-waves) we record are generally decoupled from the compressional and vertically polarised S-waves.\par
\par
\par
\par
\begin{table}[htbp]
  \caption{Acquisition parameters}
  \small
  \centering
    \begin{tabular}{|c|c|}
     \hline
    \textbf{Parameter} & \textbf{Value} \\
    \hline
    Number of source positions & 151 \\
    \hline
    Source spacing  & 2 m  \\
    \hline
    First source position  & 150.5\\
    \hline
    Last source position   & 450.5 \\
    \hline
    Number of receiver positions per source & 601 \\
    \hline
    Receiver spacing & 1 m \\
    \hline
    First receiver position & 0 m \\
    \hline
    Frequency range of the vibrator sweep & 8-250 Hz \\
    \hline
  \end{tabular}
  \label{tab:5_1}
\end{table}
\par

\newpage
\section{Results}
\begin{figure}[ht]
    \centering
    \includegraphics[width=\textwidth]{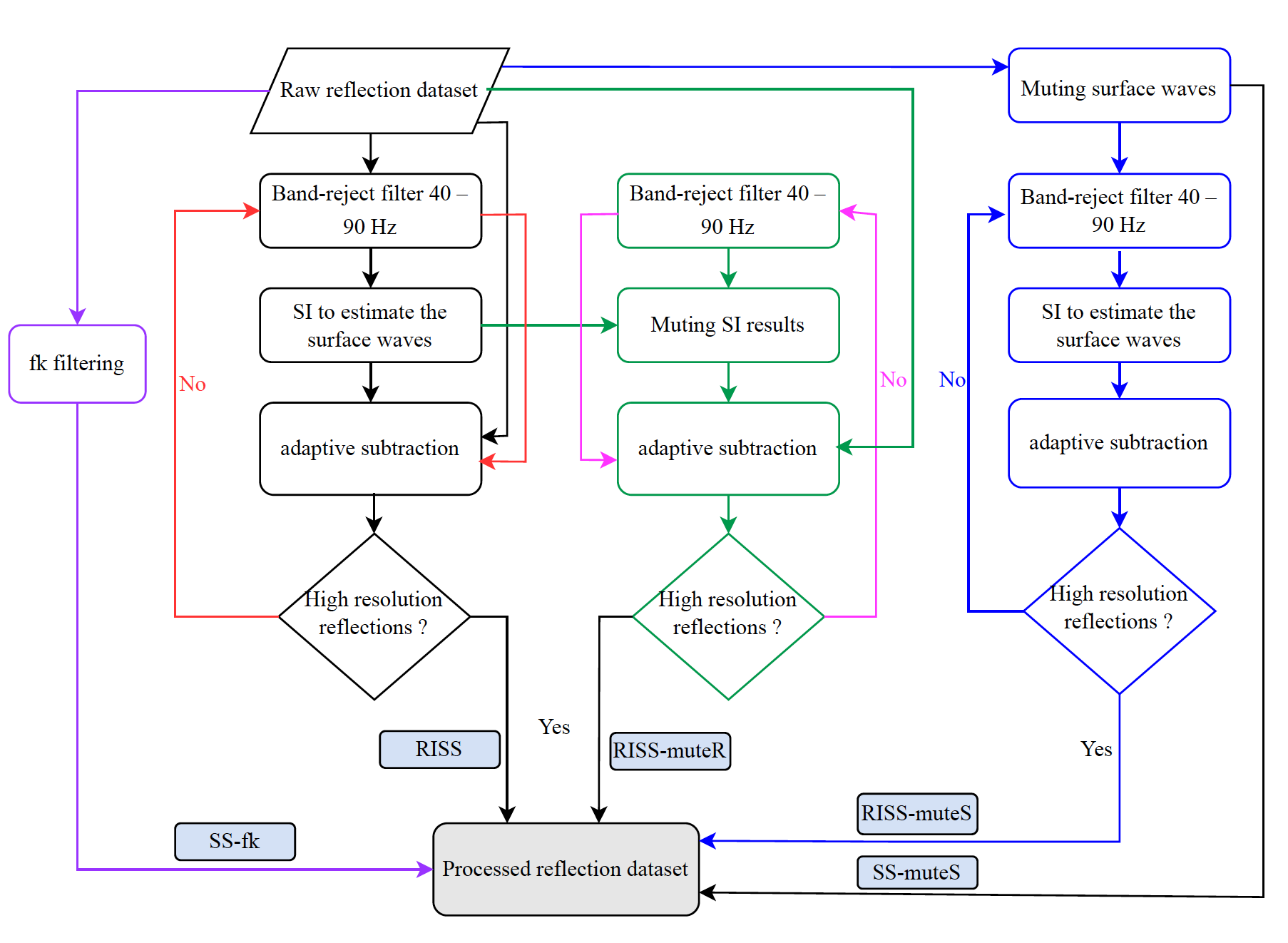} 
    \caption{Flowchart for applying surface-wave suppression. RISS stands for the Recursive Interferometric Surface-wave Suppression, the RISS-muteR is same as RISS but using muted SI results, the RISS-muteS is the same as RISS but using time-muted reflection dataset as input for SI, SS-fk denotes surface-wave suppression using f-k filtering, SS-muteS stands for surface-wave suppression using time-muted reflection dataset.}
    \label{fig:5_4}
\end{figure}

The primary aim of this study is to examine the effectiveness of the RISS technique. We apply the technique to common-source gathers of the field data, as introduced in Section 3. The data processing involves several steps. Figure~\ref{fig:5_4} shows a flowchart of data processing for RISS, but also other techniques such as f-k filtering and surgical muting for surface-wave suppression. \par

As shown in Figure~\ref{fig:5_4}, first we apply a band-reject filter between 40 Hz and 90 Hz to all active common-source gathers to reject frequencies that might contain reflections but not surface waves so that the SI result would predominantly contain retrieved surface waves. We select these frequencies based on the power spectrum of the common-source gathers. We then apply SI as described in section 2.1 by selecting each receiver close to an active source to turn it into a virtual source, as shown by the orange circles in Figure~\ref{fig:5_3}b. Next, we adaptively subtract virtual common-source gather from the original active common-source gather which is closest to the virtual source to suppress surface waves. We apply adaptive subtraction by using an estimated matching filter as described in section 2.2. \par
\begin{figure}[H]
    \centering
    \includegraphics[width=0.95\textwidth]{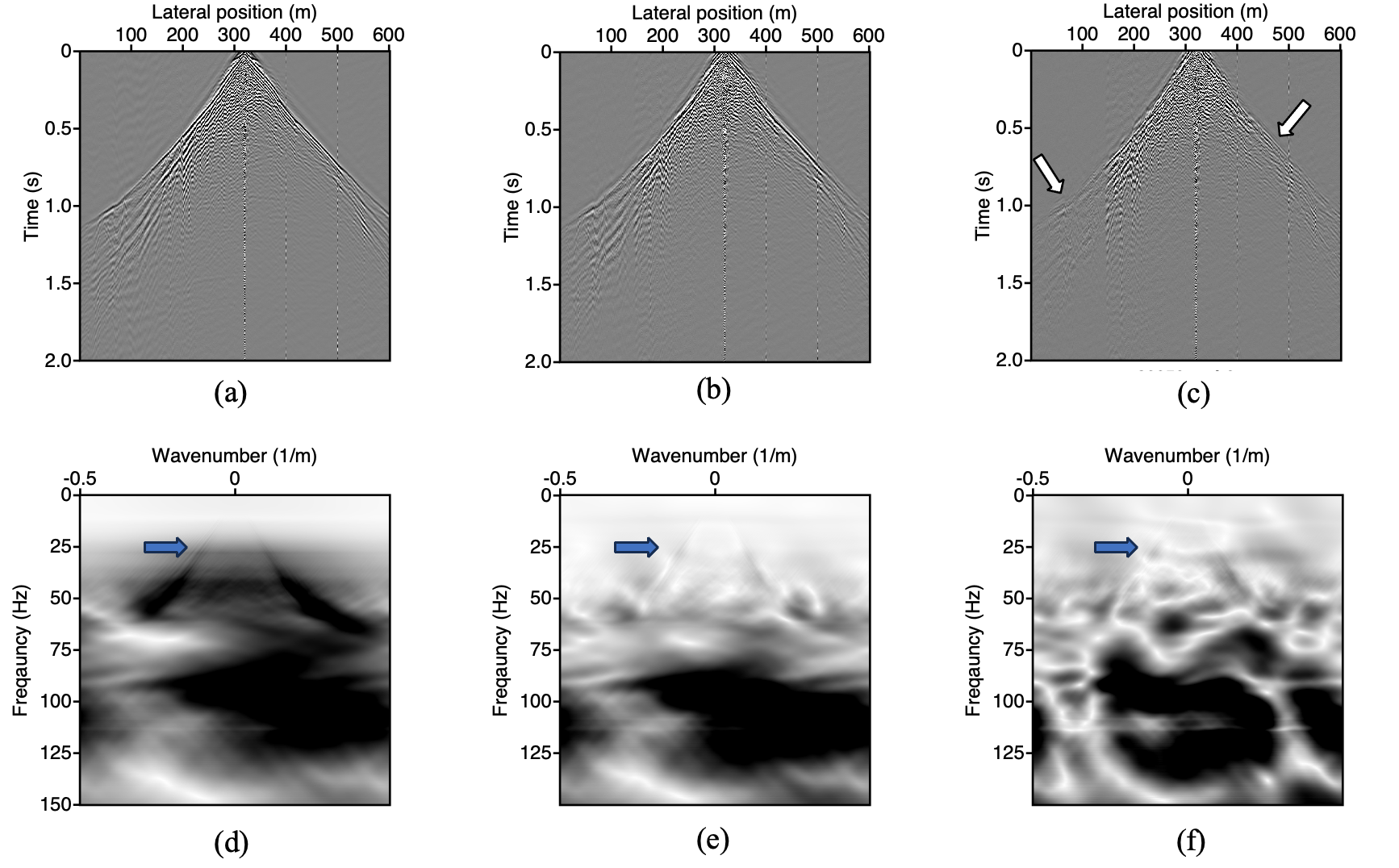} 
    \caption{(a) A common-source gather for a source located at lateral position 320.5 m, (b) same common-source gather after the first iteration of the RISS, (c) same common-source gather after the second iteration of the RISS; (d), (e) and (f) same as (a), (b), and (c), respectively, but in the f-k domain. White arrows point to suppressed surface waves in the space-time domain and blue arrows point to surface-wave energy in the f-k domain.}
    \label{fig:5_5}
\end{figure}

 
It is essential to determine carefully the key parameters for the matching filter: time window, space window, and filter length. We choose 20 traces for the space window and 0.2 s for the time window, with a filter length of 0.05 s. We apply the same steps for all virtual sources. Figure~\ref{fig:5_5}a shows an example of the original common-source gather in the time domain, while Figure~\ref{fig:5_5}b shows the same gather after RISS with one iteration, for an active source located at lateral position 320.5 m. \par 
As discussed in Section 2, we propose to apply the RISS for more than one iteration, which means we use the data after the first iteration of RISS as input for applying SI. Then, we repeat all steps i.e., band-reject filtering, SI, and adaptive subtraction. Note that these steps are shown as "RISS" in the flowchart in Figure~\ref{fig:5_4}.\par
Figure~\ref{fig:5_5}c shows the same gather as in Figures~\ref{fig:5_5}a and \ref{fig:5_5}b but after the second iteration of RISS. Comparing these three results, we observe that a large portion of the surface-wave energy is suppressed in Figure~\ref{fig:5_5}c, as indicated by the white arrows. Figures~\ref{fig:5_5}d-\ref{fig:5_5}f show the f-k spectra of the common-source gathers illustrated in Figures~\ref{fig:5_5}a-\ref{fig:5_5}c, respectively. The surface-wave energy appears as linear events in the f-k domain, as indicated by the blue arrows; they are largely suppressed from the data after the RISS with two iterations, as can be observed in Figure~\ref{fig:5_5}f. \par
\begin{figure}[H]
    \centering
    \includegraphics[width=0.95\textwidth]{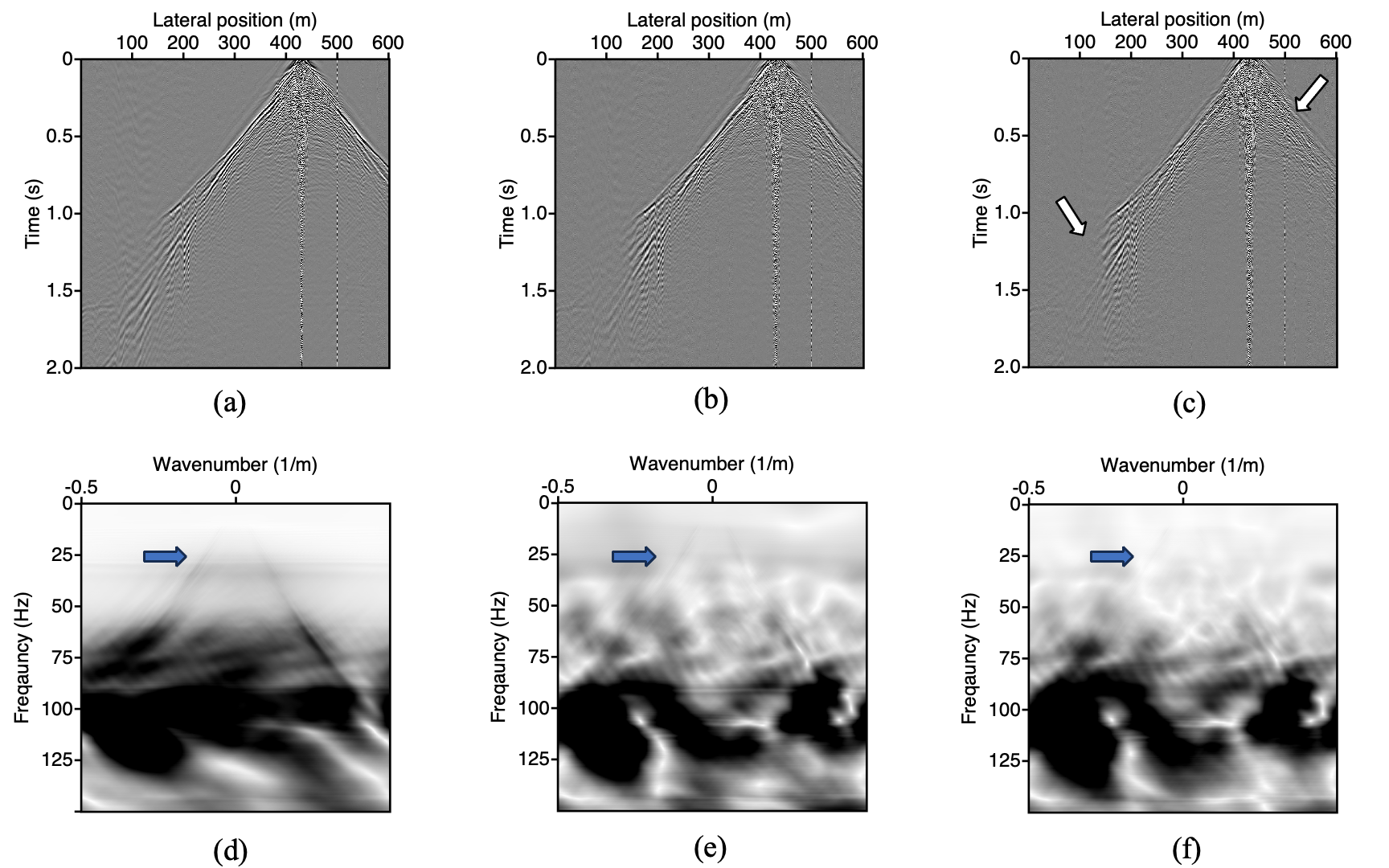} 
    \caption{Same as Figure~\ref{fig:5_5} but for a source located at a lateral position 430.5 m.}
    \label{fig:5_6}
\end{figure}

Figure~\ref{fig:5_6} shows another example for a common-source gather for a source located at a lateral position of 430.5 m, where we also observe significant suppression of the surface-wave energy in both the space-time and the f-k domains.\par

\begin{figure}[H]
    \centering
    \includegraphics[width=0.95\textwidth]{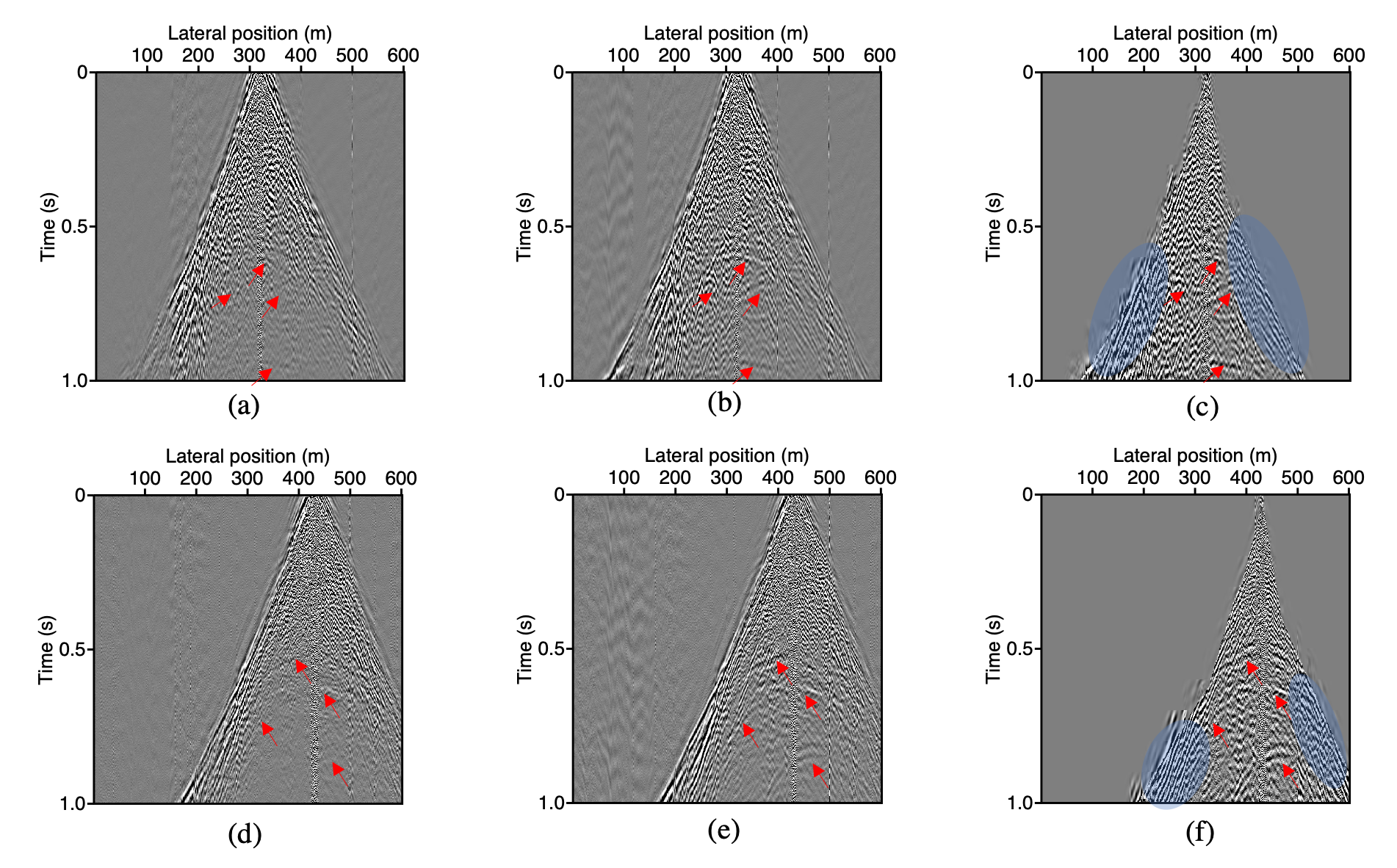} 
    \caption{(a) A common-source gather for a source located at lateral position 320.5 m after the RISS, (b) same common-source gather after the RISS-muteR, (c) same common-source gather after the RISS-muteS, (d), (e) and (f) same as (a), (b), and (c), respectively, but for a source located at a lateral position of 430.5 m. Red arrows point to enhanced reflections, while the blue ellipses highlight parts of the surface waves.}
    \label{fig:5_7}
\end{figure}

\newpage
As depicted in Figures~\ref{fig:5_5}c and \ref{fig:5_6}c, we successfully suppress the surface waves. However, it appears that some deeper reflections are also suppressed in the process (the red pointers in Figures \ref{fig:5_7}a and \ref{fig:5_7}d). This shows that applying a simple band-reject filter and relying on having fewer sources in the stationary-phase regions contributing to the retrieval of reflections might not guarantee that the retrieved reflection energy is absent or even weak. To preserve these reflections in the original common-source gathers, we apply a bottom muting to the virtual-common-source gathers retrieved from SI before adaptive subtraction, which we label as "RISS-muteR" in the flowchart in Figure~\ref{fig:5_4}. Figures~\ref{fig:5_7}b and \ref{fig:5_7}e show the common-source gather after applying the RISS using the muted SI results for two active sources located at 320.5 m and 430.5 m, respectively. In comparison to Figures~\ref{fig:5_7}a and \ref{fig:5_7}d, which show the same common-source gather after RISS, we observe here clearer deeper reflections as marked by the red arrows. \par
By examining the common-source gathers, we observe that it is feasible to suppress some parts of the surface waves through surgical muting, which is a common approach. Therefore, prior to applying the RISS, we can also surgically mute the prominent surface waves. We label this procedure as "RISS-muteS" in the flowchart in Figure~\ref{fig:5_4}. Figures~\ref{fig:5_7}c and \ref{fig:5_7}f show the common-source gather after the RISS-muteS. Although we enhance the resolution of some reflections, we still seem to have some strong surface-wave energy in comparison with the results in Figures~\ref{fig:5_7}a and \ref{fig:5_7}d as highlighted by the blue ellipses. This observation underscores the fact that by suppressing the surface waves in common-source gathers before the RISS, it becomes challenging to retrieve the surface waves by SI. \par

After the suppression of the surface waves, we apply conventional seismic processing to obtain preliminary unmigrated stacked sections for a better comparison between the result of the RISS and the conventional suppression techniques. A summary of these steps is presented in Table~\ref{tab:5_2}. We first apply bandpass filtering, and amplitude correction to compensate for intrinsic attenuation and geometrical spreading, and then automatic gain control (AGC) for visualisation purposes.\par
Next, we sort the data into common-midpoint (CMP) gathers (CMP spacing 0.5 m). As expected, the CMP fold increases towards the center of the line where better illumination is achieved. We then apply normal moveout (NMO) correction using a constant velocity of 350 m/s, and finally we stack the CMP gathers. The constant velocity of 350 m/s is selected based on a comparison of the stacked sections with different velocity values, as 350 m/s velocity yields comparatively better results. \par
\newpage
\begin{table}[htbp]
    \caption{Summary of seismic processing steps.}
     \centering
     \small
    \begin{tabular}{|c|l|}
    \hline
    \textbf{Step} & \textbf{Instruction} \\
    \hline
    1 & Band-pass filtering 30-100 Hz \\[0.1pt]
    2 & Amplitude corrections \\[0.1pt]
    3 & Automatic gain control (AGC)  \\[0.1pt]
    4 & Time muting   \\[0.1pt]
    5 & Normal moveout (NMO) correction  \\[0.1pt]
    6 & Common midpoint/ensemble stack \\[0.1pt]
    \hline
  \end{tabular}
  \label{tab:5_2}
\end{table}

\begin{figure}[H]
    \centering
    \includegraphics[width=\textwidth]{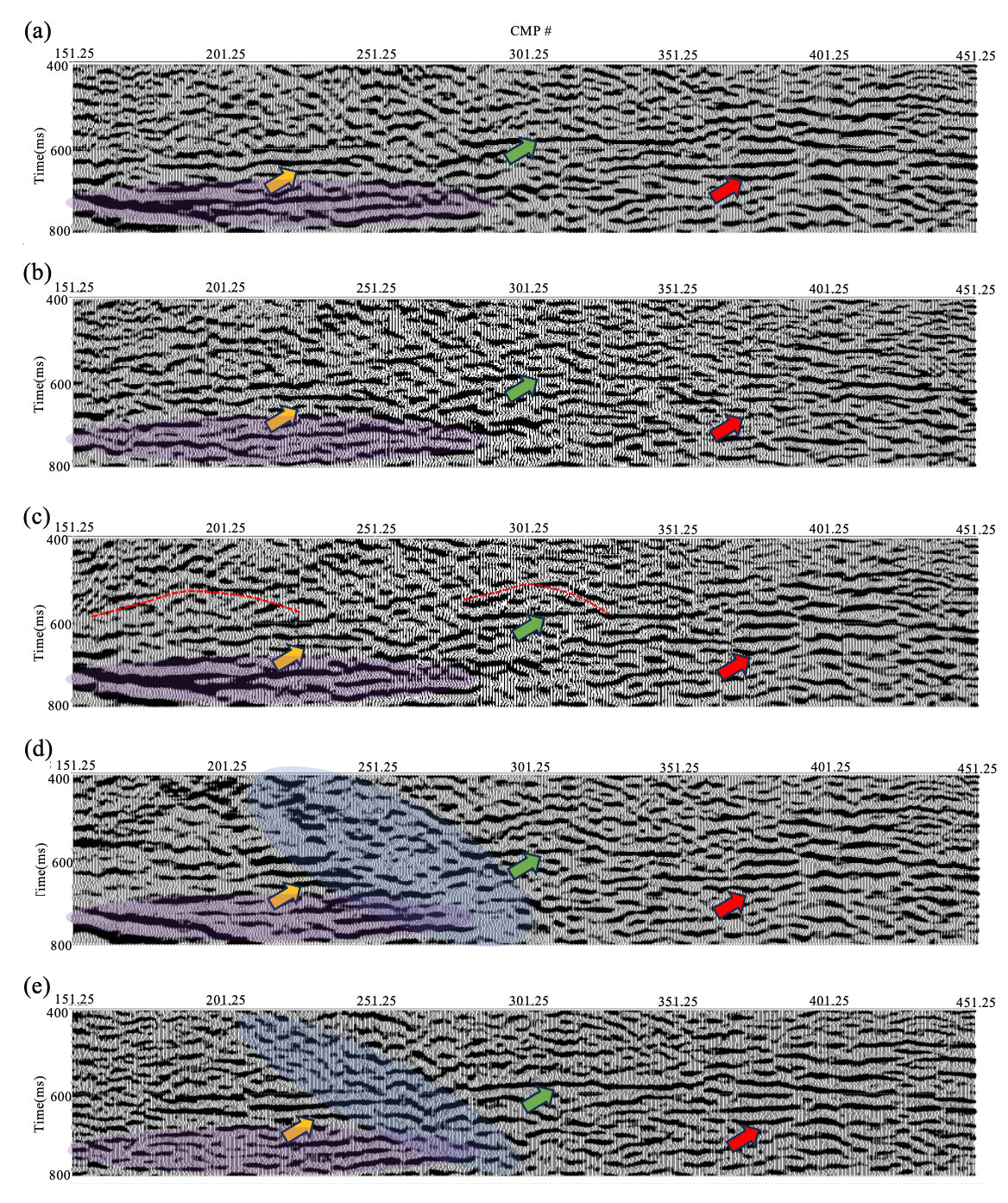} 
    \caption{Preliminary unmigrated stacked section (400-800 ms) using a constant velocity of 350 m/s: (a) using the reflection dataset after surgical muting of surface waves ("SS-muteS"), (b) after the second iteration of the RISS, (c) after the RISS using the muted SI results (RISS-muteR), (d) using the data as in (a) but after the RISS ("RISS-muteS"), (e) after f-k filtering ("SS-fk"). The colored arrows point to strong reflectors, the purple ellipses highlight the specific reflectors which are discussed in the text and the blue ellipses highlight artefacts.}
    \label{fig:5_8}
\end{figure}

Figure~\ref{fig:5_8} shows the preliminary unmigrated stacked section between 151.25 m and 450.25 m lateral distances for the five approaches to surface-wave suppression as illustrated in Figure~\ref{fig:5_4}: (a) Surgical muting (SS-muteS), (b) RISS with two iterations, (c) the RISS-muteR (same as RISS but using muted SI results), (d) surface-wave suppression using time-muted reflection dataset (RISS-muteS), and (e) surface-wave suppression using f-k filtering (SS-fk). Since we know that the most significant influence of the surface waves is related to the shallowest part of the subsurface, we focus our attention on these parts, specifically 400-800 ms. \par 
Figure~\ref{fig:5_8}a shows the time section obtained after suppressing the surface waves using surgical muting, as indicated by SS-muteS in the flowchart. We use this figure as a reference because this suppression approach is standard and experienced data processors generally achieve good results. This result is comparable with results from other studies, e.g., that in \textcite{Kruiver2017}. \par
Figure~\ref{fig:5_8}b shows the time section obtained after the second iteration of the RISS. Compared to the reference time section in Figure~\ref{fig:5_8}a, surface suppression is similar between the two sections. However, it is evident that some expected reflectors are suppressed, as indicated by the red and green arrows, as well as the reflectors within the purple ellipse. This observation highlights that using a simple band-reject filter may not guarantee that the retrieved reflection energy is either absent or significantly weakened in the results of SI.\par
To address this issue, we use the RISS-muteR as explained above. Figure~\ref{fig:5_8}c illustrates the time section obtained after the application of this technique. In comparison with the reference in Figure~\ref{fig:5_8}a, we retrieve all reflectors, as indicated by the orange, red, and green arrows. Moreover, the reflectors in the purple ellipse are preserved completely, and some dome-like structures are now interpretable as highlighted by the red curves. This improvement is due to suppressing those parts of surface waves which cannot be suppressed just by using surgical muting.  \par
Figure~\ref{fig:5_8}d shows the time section after applying RISS-muteS. Comparing this image to the images in Figures~\ref{fig:5_8}a and~\ref{fig:5_8}c, we see that the lateral continuity of the reflectors is worse, e.g., particularly inside the blue ellipse and the purple ellipse, while the general character on the left part of the image has changed, the wavefield becoming enriched with lower frequencies, which might point to the presence of left-over dominant surface-wave energy. \par
Figure~\ref{fig:5_8}e shows the time section obtained after suppression of the surface waves by f-k filtering (SS-fk). f-k filtering is commonly used for surface-wave suppression. Comparing the image in Figures~\ref{fig:5_8}e with the images in Figures~\ref{fig:5_8}a and~\ref{fig:5_8}c, we see that the result in Figure~\ref{fig:5_8}e is generally of good quality. The reflector indicated by the green arrow appears laterally more continuous than in Figures~\ref{fig:5_8}a and~\ref{fig:5_8}c. However, other reflectors are less clear, e.g., the one indicated by the purple ellipse or those in the left part of the image earlier than 600 ms (see specifically inside the blue ellipse).\par
Based on this comparison, it appears that the best image is the one presented in Figure~\ref{fig:5_8}c, i.e., the image after the application of the RISS-muteR. This is due to its clarity and its ability to provide the same information (as the event within the purple ellipse), while also offering additional details compared to the reference, such as the events shown by the red curves in the time section. Moreover, RISS is data-driven.\par

\newpage
To demonstrate the applicability of using SI for surface-wave suppression in imaging the shallow subsurface, we further investigate its use for the Marchenko-based isolation method (\cite{Wapenaar2021EMPLOYINGMETHOD}; \cite{VanIjsseldijk2023ExtractingIsolation}). The Marchenko method can estimate Green's functions between the Earth's surface and arbitrary locations in the subsurface. These Green's functions are used to redatum wavefields to a deeper level in the subsurface (\cite{Slob2013SeismicEquations}; \cite{Wapenaar2014MarchenkoImaging}). The Marchenko method can also be used to isolate the response of a specific layer or package of layers, free from the influence of the overburden and the underburden. The complete derivation of Marchenko-based isolation is beyond the scope of this paper; however, a detailed description can be found in \textcite{VanIjsseldijk2023ExtractingIsolation}, \textcite{IJsseldijkTrollb}. Moreover, a complete description of the application of Marchenko-based isolation on the field dataset used in this study can be found in \textcite{FaezehThesis}. \par
\begin{figure}[H]
    \centering
    \includegraphics[width=\textwidth]{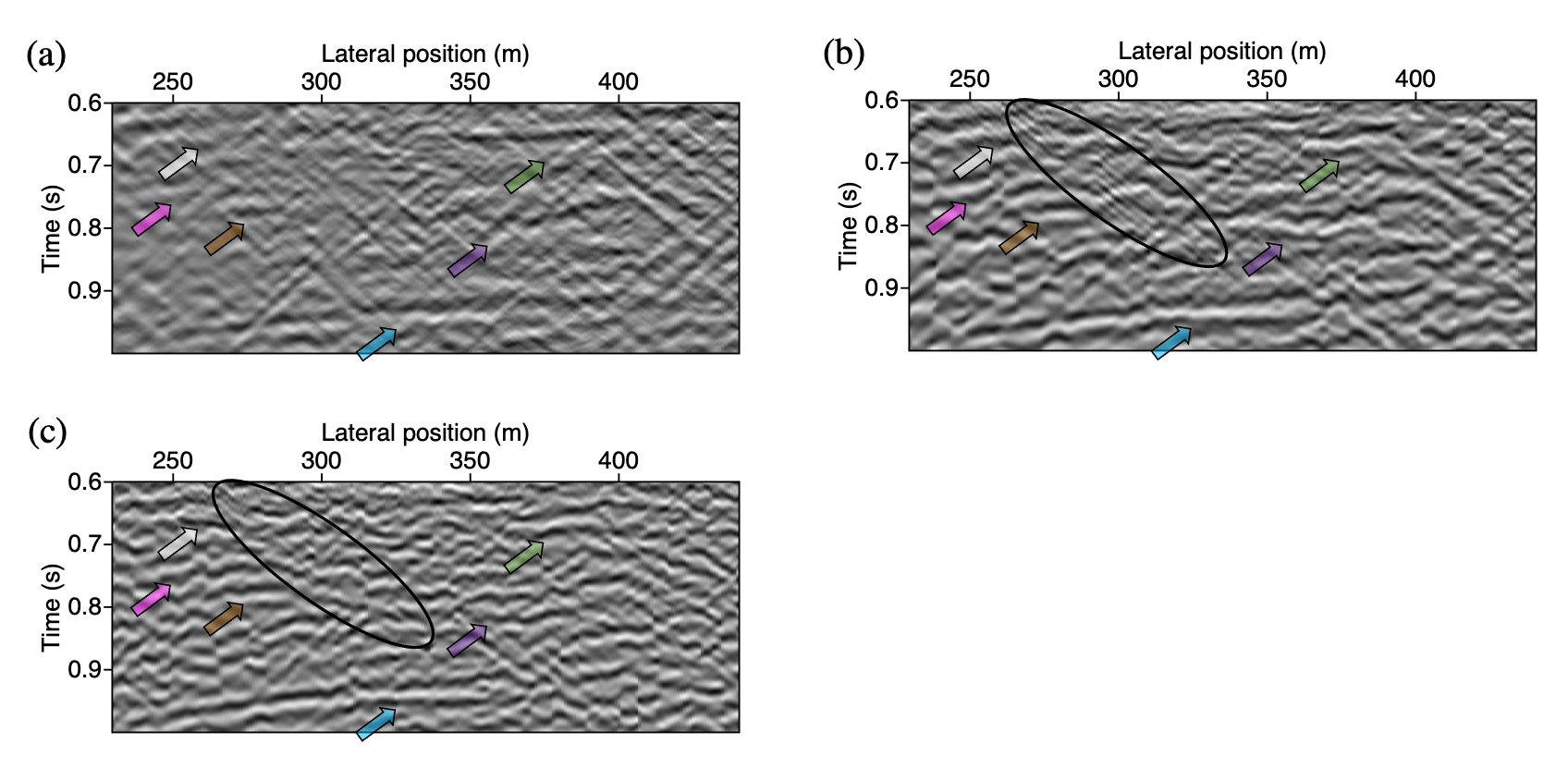} 
    \caption{Stacked sections, zoomed in between 0.6 s and 1.0 s, obtained using (a) the regular reflection response after surgical muting of the surface waves, (b) the reflection response after Marchenko-based isolation for overburden and underburden removal after surgical muting of the surface waves, and (c) similar to (b) but using data after surface-wave suppression with RISS-muteR instead of surgical muting. The colour-coded arrows indicate reflectors. The black ellipse highlights potential artefacts from the surface waves that are suppressed in (c).}
    \label{fig:5_9}
\end{figure}

 
We use the Marchenko-based isolation method to eliminate the overburden and the underburden, and isolate the reflection from the target layer between 30 m and 270 m using the data after surgical muting of surface waves and the data after RISS-muteR. Figure~\ref{fig:5_9}a shows the stacked section using the regular reflection response after suppression of the surface waves using surgical muting, while Figures~\ref{fig:5_9}b and~\ref{fig:5_9}c show the stacked section using the reflection response after Marchenko-based isolation for overburden and underburden removal, using surgical muting for surface-wave suppression and RISS-muteR, respectively. Note that we show the image plots of the section rather than wiggle representations in Figure~\ref{fig:5_8}.  \par
A comparison of these stacked sections in Figure~\ref{fig:5_9} suggests the potential elimination of the internal multiples originating from the overburden, down to 30 m using the Marchenko-based isolation. The shallow reflectors appear clearer and more continuous, as indicated by the color-coded arrows. But what we want to draw attention to is the effect of surface-wave suppression on these results. We observe enhanced reflections with greater clarity in the stacked sections after RISS-muteR (Figure~\ref{fig:5_9}c), as exemplified by the events indicated by the white and green arrows. Moreover, there are fewer artefacts, likely from surface waves, as indicated by the black ellipse. \par
\section{Discussion}
We presented a comparison of different approaches for surface-wave suppression applied to the land seismic dataset acquired in Scheemda, Groningen province: surgical muting, f-k filtering, the RISS, the RISS-muteR, and the RISS-muteS. \par
From a comparison of the unmigrated time sections, we found that surgical muting seems convenient for removing surface waves in our dataset. However, this method does not adequately remove the surface-wave energy, as surface waves overlap with useful reflections and scattered arrivals. Additionally, weak reflections and scattered arrivals covered by surface waves might also be muted, as demonstrated earlier by \textcite{KONSTANTAKI201528}.
The other common technique for surface-wave suppression --f-k filtering-- could cause more artefacts, as it is challenging to define the correct parameters for frequency and wavenumber windows for all common-source gathers. In contrast, the RISS technique can effectively suppress surface waves without prior information, as it is data-driven. \par
By comparing three approaches for the RISS, we found that the RISS using muted deeper reflections after retrieving the surface waves with SI yields the best results. RISS with two iterations was sufficient to achieve the desired results for our dataset. However, more than two iterations can be needed for other datasets.\par
For implementation of the RISS, we must determine the time, the spacing window, and the filter length. We determined these parameters carefully by examining different values for them. \textcite{Balestrini2020ImprovedSuppression} demonstrated that changing the time and the space window size makes no significant differences. They also observed that increasing the filter length produces undesirable artefacts at earlier times. Defining a proper filter length can be indeed crucial for different datasets. Moreover, we used the same parameters for both iterations, but it might be useful to change these parameters for each iteration.  \par
As described in the methodology section, we utilised the field dataset with surface waves for each active source as a reference, and for the virtual sources, the surface waves were retrieved using SI. In our study, we created the virtual sources at the receiver locations, which are positioned at 0.5 m from the active sources; this provides an appropriate dataset for applying this technique. However, using virtual sources located at greater distances from the active sources may result in incorrect estimations of the surface waves using SI. Therefore, we recommend using sources and receivers in close proximity of each other to ensure the accuracy of RISS. \par
In general, other studies, such as \textcite{Halliday2008}, have demonstrated that successfully recovering the higher-mode surface waves using only surface sources is challenging. Consequently, modal separation may be a crucial step before applying SI to ensure accurate kinematic retrieval and effective suppression of higher modes with minimal error. RISS, however, can facilitate the suppression of higher modes. During each iteration, the strongest surface-wave mode is retrieved and adaptively suppressed, effectively functioning as step-wise modal separation as discussed in Appendix A by using a numerically modelled dataset. \par

\section{Conclusion}
We proposed a recursive application of seismic interferometry (SI) for surface-wave suppression. We showcased our technique using a 2D reflection dataset acquired in Scheemda, Groningen province, the Netherlands. We applied SI to retrieve dominant surface waves between receivers while minimising the retrieved reflection energy. The retrieved dominant surface waves are then adaptively subtracted from the original data. We showed that applying these two steps two times, i.e., recursively, resulted in a fully data-driven effective suppression of the surface waves. \par 
We compared stacked sections obtained through the recursive interferometric surface wave suppression (RISS) with stacked sections where the surface waves were suppressed using f-k filtering and surgical muting. We found that the obtained time section after the second iteration of recursive interferometric surface-wave suppression yielded better results in terms of clearer and more continuous reflections, especially when the SI result was used in which the reflection energy was minimised by bottom muting before SI. This approach can be effective for enhancing the resolution of the seismic reflection events for subsurface investigations. \par

\section*{Acknowledgement}
We acknowledge the use of computational resources provided by the DelftBlue supercomputer at the Delft High Performance Computing Centre (\url{https://www.tudelft.nl/dhpc}) and thank Seismic Mechatronics for allowing us to use their vibrator source during fieldwork (\url{https://seismic-mechatronics.com/}). \par
\noindent This research is funded by NWO Science domain (NWO-ENW), project DEEP.NL.2018.048 and the reflection dataset was acquired by funding from the European Research Council (ERC) under the European Union’s Horizon 2020 research and innovation programme (Grant Agreement No. 742703) and  OYO Corporation, Japan: OYO-TUD Research Collaboration (Project code C25B74).

\section*{Data availability}
The field reflection dataset used in this study is available in the 4TU.ResearchData repository at \url{https://doi.org/10.4121/a8553b7e-82ae-4e9b-bc54-2a6b9ca6063c}.\\The codes are also accessible in the 4TU.ResearchData repository at \url{https://doi.org/10.4121/6271c7d3-f931-49e9-b2b0-2d05eef7d3ae}.

\section*{Appendix A: Suppression of different modes of surface waves using a numerically modelled dataset}
To investigate the possibility of suppressing higher modes of surface waves using RISS, we use numerical modelling based on a velocity model in Figure 3 of \textcite{Halliday2008} which has been summarised in Table~\ref{tab:A1}. We use a finite-difference modelling code (\cite{Thorbecke2011Finite-differenceInterferometry}) in an acoustic mode with a rigid boundary at the surface to generate the full seismic dataset, including reflections and surface waves. Based on the analogy between 2D acoustic waves in a fluid and 2D SH waves in a solid (\cite{Wapenaar2001}), this approach is equivalent to generating SH waves including Love waves, in the elastic mode with a free surface boundary at the surface. The fixed receivers are placed from 100 m to 400 m with 1 m spacing, and the sources are placed from 50.25 m to 450.25 m with 2 m spacing. \par

\setcounter{table}{0} 
\renewcommand{\thetable}{A\arabic{table}}

\begin{table}[htbp]
    \caption{Parameters of the velocity model used for numerical modelling (\cite{Halliday2008})}
     \centering
     \small
     \begin{tabular}{|c|c|c|}
    \hline
    \textbf{Depth (m)} & \textbf{\( V_s \) (m/s)} & \textbf{Density (kg/m\(^3\))} \\ \hline
    0-1 & 101 & 1400 \\ \hline
    1-2 & 126 & 1460 \\ \hline
    2-4 & 127 & 1470 \\ \hline
    4-6 & 146 & 1520 \\ \hline
    6-8 & 172 & 1590 \\ \hline
    8-12 & 184 & 1610 \\ \hline
    12-20 & 200 & 1650 \\ \hline
    20-30 & 232 & 1710 \\ \hline
    30- & 307 & 1840 \\ \hline
  \end{tabular}
  \label{tab:A1}
\end{table}

Figure~\ref{fig:5_A1}a shows a common-source gather for a source at 100.25 m in the time domain, and Figure~\ref{fig:5_A2}a shows the same common-source gather in the f-k domain. In this figure, the fundamental mode of surface waves is indicated by a blue arrow, and the first higher mode is indicated by a purple arrow. \par
We apply RISS to this dataset. Figure~\ref{fig:5_A1}b shows the common-source gather after the first iteration, and Figure~\ref{fig:5_A1}c shows the common-source gather after the second iteration. Figures~\ref{fig:5_A2}b and~\ref{fig:5_A2}c also show the same common-source gather in the f-k domain after the first and second iterations, respectively. To better evaluate the suppression of different modes of surface waves, the results of seismic interferometry before adaptive subtraction are shown in Figures~\ref{fig:5_A2}d and ~\ref{fig:5_A2}e, after the first and second iterations, respectively.\par
As we can see in Figures~\ref{fig:5_A1}b and~\ref{fig:5_A1}c, the strong surface waves have been effectively suppressed, and the resolution of the reflections has increased, as indicated by the red arrows. However, it is challenging to discriminate between the fundamental mode and the higher modes of surface waves in the time-space domain. Nevertheless, it is clear from the common source gathers in the f-k domain that we effectively estimated the fundamental mode of surface waves in the result of SI after the first iteration as shown in Figure ~\ref{fig:5_A2}d. As a result of the subtraction, we mainly removed the fundamental mode in the first iteration, but higher modes of the surface waves still remained (Figure~\ref{fig:5_A2}b). In the second iteration, we further estimated the remaining part of the fundamental mode and the higher mode as shown in the result of SI in Figure~\ref{fig:5_A2}e. As a result of the subtraction, we suppressed a large part of the fundamental mode and also the first higher mode of surface waves as shown in Figure\ref{fig:5_A2}c. \par
In conclusion, we demonstrated that during the two iterations, the strongest part of the surface-wave mode is retrieved and adaptively suppressed, effectively achieving step-wise modal separation. This provides the possibility of suppressing the fundamental mode and the higher modes of surface waves. It is important to note that in this example, we used a simple subsurface model with a clear separation between surface wave modes. However, in the field data set, it is not as straightforward to determine which part of the surface waves is eliminated in each iteration. Nonetheless, based on this numerical modelling, it is reasonable to assume that the iterative procedure can suppress the fundamental modes and (some of) the higher-order modes. \par

\setcounter{figure}{0} 
\renewcommand{\thefigure}{A\arabic{figure}}

\begin{figure}[h]
    \centering
    \includegraphics[width=\textwidth]{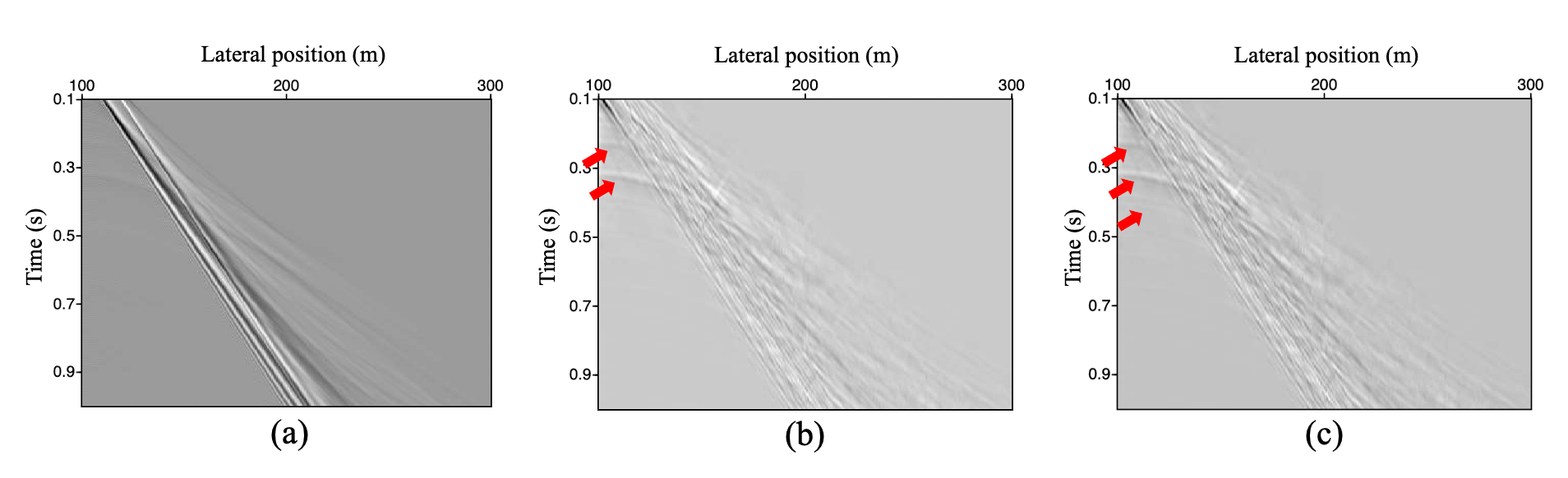} 
    \caption{A common-source gather for a source located at lateral position 100.25 m, (b) same common-source gather after the first iteration of the RISS, (c) same common-source gather after the second iteration of the RISS.} 
    \label{fig:5_A1}
\end{figure}

\begin{figure}[h]
    \centering
    \includegraphics[width=\textwidth]{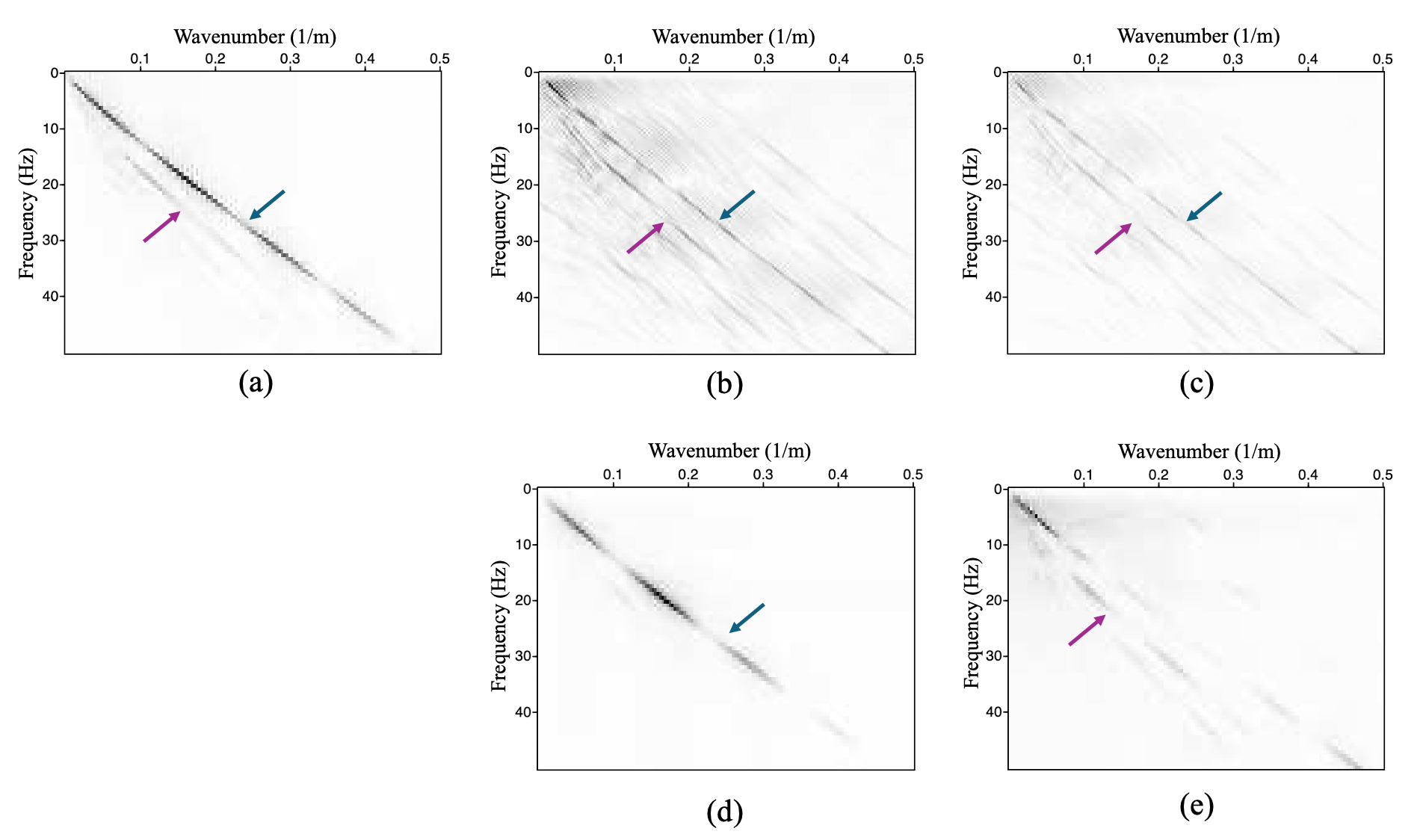} 
    \caption{A common-source gather for a source located at lateral position 100.25 m, in the f-k domain, (b) same common-source gather after the first iteration of the RISS, (c) same common-source gather after the second iteration of the RISS, (d) virtual common-source gather (i.e., result of SI before adaptive subtraction) at lateral position 100 m in the f-k domain after the first iteration, and (e) same as (d) after the second iteration.} 
    \label{fig:5_A2}
\end{figure}

\newpage
\printbibliography

\end{document}